\documentclass{PoS}

\newcommand{\dd}{{\mathrm{d}}}
\newcommand{\Z}{{\mathbf{Z}}}

\def\di{\displaystyle}

\def\bg{\begin{eqnarray}\begin{array}{rcl}\displaystyle}
\def\eg{\end{array} &\di    &\di   \end{eqnarray}}
\def\bm#1{\begin{eqnarray}\begin{array}{#1}\di}
\def\bmo#1{\begin{eqnarray*}\begin{array}{#1}\di}
\def\bml#1#2{\begin{eqnarray}\begin{array}{#1}\label{#2}\di}
\def\bgo{\begin{eqnarray*}\begin{array}{rcl}\displaystyle}
\def\ego{\end{array} &\di    &\di \nonumber  \end{eqnarray*}}

\def\btensor#1#2{\left#1\begin{array}{#2}\di}
\def\brtensor#1#2#3{\ren#3\left#1\begin{array}{#2}}
\def\botensor#1#2{\renew\left#1\begin{array}{#2}}
\def\etensor#1{\end{array}\right#1}

\def\eq#1{(\ref{#1})}
\def\Eq#1{Eq.~(\ref{#1})}

\def\tr{{\rm tr}}

\def\id{1\!\mbox{l}}

\def\s0#1#2{\mbox{\small{$ \frac{#1}{#2} $}}}
\def\0#1#2{\frac{#1}{#2}}

\def\CB{{\mathcal B}}
\def\CC{{\mathcal C}}
\def\CD{{\mathcal D}}

\def\CP{{\mathcal P}}

\newcommand{\nn}{\nonumber}

\title{On Majorana fermions on the lattice}

\ShortTitle{On Majorana fermions on the lattice}

\author{\speaker{Jan M. Pawlowski}%
	 \\
        Institut f\"ur Theoretische Physik, Universit\"at Heidelberg,
	Philosophenweg 16,\\
	69120 Heidelberg, Germany\\
        E-mail: \email{j.pawlowski@thphys.uni-heidelberg.de}}

\author{Yuji Igarashi\\
Faculty\ of\ Education, 
 Niigata\ University, Ikarashi,\\
950-2184, Niigata, Japan\\ 
E-mail: \email{igarashi@ed.niigata-u.ac.jp} }

\abstract{
  The construction of massless Majorana fermions with chiral Yukawa 
  couplings on the lattice is considered. We find topological obstructions 
  tightly linked to those underlying the Nielsen-Ninomiya no-go theorem. In 
  contradistinction to chiral fermions the obstructions originate only 
  from the combination of the Dirac action and the Yukawa term. These findings 
  are used to construct a chirally invariant lattice action. We also show 
  that the path intgral of this theory is given by the Pfaffian of the 
  corresponding Dirac operator. }

\FullConference{The XXV International Symposium on Lattice Field Theory\\
		 July 30-4 August 2007\\
		 Regensburg, Germany}

\begin{document}

\section{Introduction}

Massive neutrinos can be incorporated in the Standard model with
Majorana fermions that become massive via spontaneous symmetry
breaking. This mass generation relates to a chirally symmetric Yukawa
term. Majorana fermions with chiral symmetry also play a r$\hat{\rm
  o}$le for physics beyond the standard model, e.g.\ in supersymmetric
theories. A lattice approach to the related physics problems has to be
based on an appropriate lattice formulation of Majorana fermions in
the presence of chiral symmetry
\cite{Ginsparg:1981bj,Neuberger:1997fp,Hasenfratz:1998ri,Luscher:1998pq},
for Majorana fermions on the lattice see e.g.\ 
\cite{Montvay:2001ry,Fujikawa:2002fi,Suzuki:2004ht,Giedt:2007qg,Wolff:2007ip}.

In the present note we want to address the related obstructions and
provide a construction of chirally coupled Majorana fermions. We
believe that this construction may also prove useful for the
construction of supersymmetric theories on the lattice. Within a
lattice formulation, chiral symmetry becomes non-trivial due to the
Nielsen-Ninomiya no-go theorem
\cite{Karsten:1980wd,Nielsen:1980rz,Karsten:1981gd,Friedan:nk} , and
we expect related obstructions for chirally coupled Majorana fermions.
Indeed there appears a certain conflict between the definition of the
Majorana fermions and lattice chiral symmetry in the presence of
Yukawa couplings.  The conflict is closely related to the requirements
of locality and of avoiding species doubling, which are the basic
issues of lattice chiral symmetry. It causes an obstruction in
constructing the simplest supersymmetric model, the Wess-Zumino model
on a lattice, and also in showing CP invariance of chiral gauge
theory, see e.g.\cite{Fujikawa:2002vj,Hasenfratz:2005ch}.

Before we discuss the lattice obstruction we want to recapitulate the
continuum formulation of chirally coupled Majorana fermions with an
emphasis on 4-dim Euclidean space-time. Majorana fermions are neutral
fermions and obey a reality constraint. In 4-dim Eulicdean space-time
the charge conjugation operator $C$ has the properties
\begin{eqnarray}\label{eq:charge} 
C \gamma_\mu C^{-1}=-\gamma_\mu^T\,, \qquad 
C \gamma_5 C^{-1}=\gamma_5^T\,, \qquad C^\dagger C=\id 
\, \,, \qquad C^T= -C\,. 
\end{eqnarray} 
Majorana fermions are defined via the reality constraint
\begin{eqnarray}\label{eq:reality} 
\psi^*=B\psi\,,
\end{eqnarray}
where $C=B\gamma_5$. However, \eq{eq:charge} implies $B^* B=-\id$ and
hence we cannot implement the reality constraint \eq{eq:reality} as it
fails to satisfy the consistency condition $\psi^{**}=\psi$.  Doubling
the degrees of freedom suffices to implement the reality constraint
with
\begin{eqnarray}\label{eq:symplecticreality} 
  \psi^*= \CB \psi \,, \qquad {\rm with} \qquad \psi=\btensor{(}{c} \psi_1 \\ 
  \psi_2 \etensor{)}\,,  
  \qquad {\rm and} \qquad \CB= \btensor{(}{ccc} 0 & & B \\ -B & &  0 \etensor{)} 
  \,.
\end{eqnarray} 
The symplectic structure of $\CB$ leads to $\CB^*\CB=\id$ following
from \eq{eq:charge} with $B^* B=-\id$. Thus the reality constraint,
$\psi^{**}=\psi$, is satisfied. The corresponding charge conjugation
operator is provided by
\begin{eqnarray}\label{eq:Ccharge} 
  \CC= \btensor{(}{ccc} 0 & & C \\ C & &  0 \etensor{)} = \CB \Gamma_5\,,\qquad 
  {\rm with}\qquad \Gamma_5=\btensor{(}{ccc} -\gamma_5 & & 0 \\ 0 & &  \gamma_5 
  \etensor{)}\,.
\end{eqnarray} 
The above properties of the symplectic Majorana fermion $\psi$ fix its
behaviour under chiral rotations,
\begin{eqnarray}\label{eq:chiral} 
  \psi\to (1+ i \epsilon\Gamma_5)\psi\,. 
\end{eqnarray} 
Now we are in the position to construct a chirally invariant Majorana
action.  We summarise the necessary properties,
\begin{eqnarray}\label{eq:Cprops} 
  \CC=\CB\Gamma_5\,,\qquad 
  \CC \Gamma_5\CC^{-1}=-\Gamma_5^T\,, \qquad \CC^\dagger \CC=\id 
  \, \,, \qquad \CC^T= -\CC\,,  
\end{eqnarray} 
and construct the corresponding chirally invariant Majorana action
\begin{eqnarray}\label{eq:majocont} 
  S[\psi]=\int d^4 x\, \psi^T \CC \CD\psi= 
  \int d^4 x\, \left(\psi_1^T C D\psi_1+\psi_2^T C D
    \psi_2\right)
\end{eqnarray} 
with
\begin{eqnarray}\label{eq:majocont1}
\CD=
  \btensor{(}{ccc} 0 & & D \\ D & &  0 \etensor{)}\,, \qquad {\rm and} 
\qquad (\CC\CD)^T=-\CC\CD \,. 
\end{eqnarray} 
The action \eq{eq:majocont} could also be obtained by a Majorana
reduction, see e.g. \cite{Nicolai:1978vc,vanNieuwenhuizen:1996tv}. 
We remark that skew symmetry of $\CC\CD$ is not required but only the
  skew-symmetric part of $\CC\CD$ contributes to the action $S$. The
definitions \eq{eq:majocont1} imply
\begin{eqnarray} \label{eq:skew} (C\, D)^T=-C\,D \,,\qquad {\rm and}
  \qquad D^*={B D B^{-1}\,},
\end{eqnarray} 
for the Dirac operator $D$. The combined properties \eq{eq:skew} hold
for the standard chiral Dirac operator with
\begin{eqnarray}\label{eq:schiral} 
  \gamma_5 D+D\gamma_5 =0\,,
\end{eqnarray} 
such as $D=\gamma_\mu \partial_\mu$, for which \eq{eq:skew} can be
deduced from \eq{eq:charge}. The action \eq{eq:majocont1} is chirally
invariant under a chiral transformation with \eq{eq:chiral} if
\begin{eqnarray}\label{eq:chiraldirac} 
  \Gamma_5 \CD-\CD\Gamma_5 = 0\,, 
\end{eqnarray} 
which is valid for $\CD$ with the standard Dirac operator
\eq{eq:schiral}.  Finally we remark that the action \eq{eq:majocont}
is real, as follows from \eq{eq:skew}. It is instructive to make this
reality explicit by rewriting the action \eq{eq:majocont} with the
help of the above relations,
\begin{eqnarray}\label{eq:majocontreal} 
  S[\psi]
  =\int d^4 x\, \left(\psi_2^\dagger \gamma_5 D\psi_1 +\psi_1^\dagger 
    \gamma_5 D^\dagger \psi_2\right)\,. 
\end{eqnarray} 
Note that \eq{eq:majocontreal} is even real for unconstrained Dirac
fermions $\psi_1,\psi_2$ in contrast to \eq{eq:majocont}.  For the
construction of a chirally invariant Yukawa term we introduce chiral
projection operators related to $\Gamma_5$ in \eq{eq:chiral},
\begin{eqnarray}\label{eq:P}
  \CP=\012 (1+\Gamma_5)=\btensor{(}{ccc} P & & 0 \\ 0 & &  
  1-P \etensor{)} \,,\qquad {\rm with}\qquad P= \012 (1-\gamma_5)\,.
\end{eqnarray} 
The chiral projection operators $\CP$, $(1-\CP)$ allows us to project
on left-handed and right-handed spinors. With these projections we can
couple the Majorana fermions to a chirally invariant Yukawa term,
\begin{eqnarray}\nonumber 
  S_{Y}[\psi,\phi] &= &g  
  \int d^4 x\left(\, {\psi}^{T}\CC \CP 
    \phi^{\dagger} ~  {\psi}+ {\psi}^{T}\CC(1-\CP) 
    \phi  {\psi}\right)\\
  &=&g\int d^4 x \bigl( 
  {\psi_1}^{T}C\, P 
  \varphi ~ {\psi_1}+ {\psi_1}^{T}C\, (1-P) 
  \varphi^{\dagger}  {\psi_1}\nonumber\\
  && \qquad\qquad    +{\psi_2}^{T}C\, P 
  \varphi^{\dagger} ~ {\psi_2}+ {\psi_2}^{T}C\, (1-P) 
  \varphi  {\psi_2}  \bigr) \,,  
\label{eq:S_Y}\end{eqnarray}
where $\varphi$ is a complex scalar,
\begin{eqnarray}\label{eq:phi}
  \phi= \btensor{(}{ccc} 0 & & \varphi \\ \varphi & &  0 \etensor{)}\,, \qquad 
  {\rm with} \qquad  \phi \to (1{-}2i\epsilon)\phi\,.
\end{eqnarray} 
Note that the scalar field $\phi$ is off-diagonal and hence does not
commute with the projection operators, we have e.g.\, $\CP
\phi^\dagger =\phi^\dagger (1-\CP)$. The action
$S[\psi]+S_Y[\psi,\phi]$ is invariant under the transformation
\eq{eq:chiral} of the fermions and that in \eq{eq:phi} for the scalar
field $\phi$ related to $\varphi\to (1 {-}2i\epsilon)\varphi$. This
concludes our brief summary of chirally coupled Majorana fermions in
the continuum. Due to chiral symmetry, and in particular the use of
chiral projections in \eq{eq:S_Y} we expect obstructions for putting
the above theory on the lattice.

\section{Lattice formulation and topological obstructions for Majorana
  fermions} 
Chiral symmetry on the lattice differs from that in the continuum as
consistent chiral transformations necessarily depend on the Dirac
operator.  Hence we first dicuss the properties of the lattice version
of the Dirac action \eq{eq:majocont}
\begin{eqnarray}\label{eq:majolattice}
  S[\psi] =
  \sum_{x,y\in\Lambda} \psi^T(x) \CC \CD(x-y) \psi(y)\,, 
\end{eqnarray} 
with the lattice Dirac operator $D(x-y)$ used in the definition of
$\CD$ as defined in \eq{eq:majocont1}. Assume for the moment that
$D(x-y)$ is of Ginsparg-Wilson type \cite{Ginsparg:1981bj} with
\begin{eqnarray}\label{eq:GWga} 
  \gamma_5 {D}+{D}\gamma_5= a {D}\gamma_5{D}\,. 
\end{eqnarray}
Then the chiral transformation
\begin{eqnarray}\label{eq:GWchiral}
  \psi\to (1+ i \epsilon\Gamma_5 (1-\012 a D))\psi\,, 
\end{eqnarray} 
is an invariance of \eq{eq:majolattice}. However, smooth chiral
projections $P$ and $\CP$ cannot be constructed, which is reflected in
the fact that the transformation \eq{eq:GWchiral} vanishes at the
doublers.  This is a consequence of the well-known Nielsen Ninomiya
(NN) no-go theorem
\cite{Karsten:1980wd,Nielsen:1980rz,Karsten:1981gd,Friedan:nk} , which
provides obstructions for putting chiral fermions on the lattice.
Ginsparg-Wilson fermions \cite{Ginsparg:1981bj} circumvent the no-go
theorem with a modified chiral symmetry \eq{eq:GWga}, which can be
reformulated as
\begin{eqnarray}\label{eq:GW} 
  \gamma_5 {D}+{D}\hat \gamma_5=0, \qquad {\rm with} 
  \qquad \hat \gamma_5=
  \gamma_5 (1-a {D })\,, 
\end{eqnarray} 
and chiral projections 
\begin{eqnarray}\label{eq:project} 
  P= \s012 (1 {-} \gamma_5),\qquad \hat P =
  \s012(1 {-} \hat \gamma_5)\,.
\end{eqnarray}
The general case going beyond Ginsparg-Wilson fermions, including
e.g.\cite{Fujikawa:2000my,Fujikawa:2002is}, 
only resorts to general chiral projections
$P,\hat P$, which are compatible:
 \begin{eqnarray}\label{eq:genchiral} 
   (1-P)\, D = D\, \hat P\,. 
\end{eqnarray}
It has been shown in \cite{Jahn:2002kg} that projection operators
$P,\hat P$ carry a winding number that is related to the total
chirality $\chi$ of the system at hand,
 \begin{eqnarray}\label{eq:chirality} 
   \chi  = n[\hat P]-n[1-P]\,, \qquad {\rm with} \qquad 
   n[P] \equiv \frac{1}{{2}!}\left(\frac{i}{2\pi}\right)^{2} 
   \int_{T^{4}} \tr\,P (\dd P)^{4} \,.
   \in \Z \,, 
\end{eqnarray}
if $\hat P\psi =\psi$ in the action. \Eq{eq:chirality} also entails
that for odd chirality $\chi$, $\hat P\psi$ and $P \psi$ live in
topologically different spaces, and hence have to be different. In the
present case the total chirality $\chi$ is even due to the symplectic
construction. The continuum Yukawa action, however, contains
projection operators $\CP,1-\CP$ with $P,1-P$ on chiral sub-spaces
with $\CP \psi\neq \psi$, that is on fermionic sub-systems with odd
chirality.  Thus we have to worry about the use of projection
operators in the Yukawa action $S_Y$.

The first question that arises in this context is whether the lattice
Yukawa action can be constructed such that it is left invariant under
the chiral transformations \eq{eq:GWchiral}, and tends toward the
continuum action.  This would require the existence of a smooth
operator $\tilde P$ which reduces $\tilde P\to P$ in the continuum
limit, and ensures invariance of the Yukawa term under the combined
transformation \eq{eq:GWchiral} and \eq{eq:phi}. However, as
$\gamma_5(1-\0{a}{2}D)$ is not normalised and even vanishes at the
doublers such an operator $\tilde P$ cannot exist, even if one relaxes
the projection property $\tilde P^2=\tilde P$, see also
\cite{Igarashi:2002bs}. This important results will be detailed
elsewhere.

In turn it is required that the chiral transformation must be
compatible with the projection operators used in the Yukawa term. This
already excludes \eq{eq:GWchiral}. Without loss of generality we can
restrict ourselves to the chiral transformation
\begin{eqnarray}\label{eq:chiraltrafo} 
  \psi\to (1+i \epsilon \hat\Gamma_5)\psi\,, \qquad \longrightarrow \qquad 
  \psi^T\hat\CC 
  \to \psi^T\hat\CC\,[
  \hat\CC^{-1} (1+i \epsilon \hat\Gamma_5^T) \hat\CC]\,,
\end{eqnarray} 
where $\hat\CC$ is a lattice generalisation of $\CC$.  Then, chiral
invariance of the action $S$ in \eq{eq:majolattice} leads to the
constraint
\begin{eqnarray}\label{eq:constraintG} 
  \hat\CC^{-1} \hat\Gamma_5^T \hat\CC = -\Gamma_5\,,\qquad {\rm with} \qquad 
  \Gamma_5 \CD= \CD\hat\Gamma_5\,. 
\end{eqnarray} 
We conclude that invariance of the lattice action \eq{eq:majolattice}
under the chiral transformations \eq{eq:chiraltrafo} would require
\begin{eqnarray}\label{eq:compatible}
  \hat \gamma_5^T =\hat C \gamma_5 {\hat C}^{-1} \,, 
\end{eqnarray}
which maps $\hat\gamma_5$ carrying the winding number $n[\hat P]$ to
$\gamma_5$ carrying the winding number $n[P]$. Note that using
different $\gamma_5$'s in the definition of $\Gamma_5$ still leads to
the same conclusion \eq{eq:compatible}. In order to elucidate this
obstruction we use Ginsparg-Wilson fermions as an example.  There the
relation \eq{eq:compatible} reads
\begin{eqnarray}\label{eq:compatibleGW} 
  (1-a D^T)\gamma_5^T =\hat C \gamma_5 {\hat C}^{-1} \,, 
\end{eqnarray}
with a possible solution 
\begin{eqnarray}\label{eq:solGW} 
  \hat C =C (1-\s012 a D)\,.
\end{eqnarray} 
The $\hat C$ in \eq{eq:solGW} has zeros at the doublers, and the
relative winding number is carried by these zeros. Inserting a lattice
$\hat C$ in \eq{eq:solGW} into the action \eq{eq:majolattice} we
encounter zeros or singularities of the operator ${\hat C}^{-1} D$ at
the positions of the doublers. This brings back the doubling problem.
Consequently {we} have to use {\it independent} Majorana fields
$\psi$, $\psi'$ with different chiral transformation properties for
the construction of Majorana actions.

\section{Construction of Majorana actions on the lattice}

Now we are in the position to construct chirally coupled Majorana
fermions on the lattice.  In line with the arguments of the last
section we introduce a copy of the original symplectic Majorana
fermion, $\psi'$. Then chiral invariance is easily arranged for with
appropriate, different, chiral transformations for $\psi$ and $\psi'$
respectively. Furthermore we have to ensure that our path integral
results in Pfaffians of the Dirac operator which signals Majorana
fermions. The corresponding lattice action reads
\begin{eqnarray}\label{eq:majolatticedoub}
  S[\psi,\psi'] =
  \sum_{x,y\in\Lambda} {\psi'}^T(x) \CC \CD(x-y) \psi(y)\,,  
\end{eqnarray} 
with the Yukawa term 
\begin {eqnarray}  S_{Y}[\psi,\psi',\phi] &= &g
  \sum_{x,y\in\Lambda}\left(\, {\psi'}^{T}\CC \CP ~ {\phi}^\dagger
    (1-\hat \CP) {\psi}+ {\psi'}^{T}\CC (1-\CP) {\phi} \hat\CP
    {\psi}\right)
  \label{eq:S_Ylattice}
\end{eqnarray}
where ${\cal P}=(1 +
\Gamma_{5})/2, ~\hat{\cal P}=(1+ \hat\Gamma_{5})/2$. We emphasise that 
in contradisctinction to the continuum theory in general we have $\CP
\phi^\dagger \neq \phi^\dagger (1-\CP)$ and $\hat \CP
\phi^\dagger \neq \phi^\dagger (1-\hat \CP)$ as the projection
operators $\CP,\hat \CP$ depend on the Dirac operator.
The action $S+S_Y$ is invariant under the chiral transformations
\begin{eqnarray}\label{eq:chirallat} 
  \psi\to (1+i\epsilon\hat\Gamma_5) \psi\,,\qquad 
  \psi'\to  (1+i\epsilon\Gamma_5 )\psi'\,,\qquad \phi\to (1 -
  2 i\epsilon) \phi\,.
\end{eqnarray} 
The action $S+S_Y$ reduces to the continuum action in the continuum
limit, but with a doubling of the field content. This doubling can be
removed by appropriate prefactors in the action, or by simply taking
roots of the generating functional $Z$. However, it is left to prove
the Pfaffian nature of the path integral. Since we have doubled the
degrees of freedom we could have constructed a Dirac fermion out of
two Majorana fermions. To that end we concentrate on the path integral
of the pure Majorana action \cite{Suzuki:2000ku,Inagaki:2004ar}
including a mass term for dealing with the zero modes.  The generating
functional is given by
\begin{eqnarray}\label{eq:pathmajo}
  Z=\int \prod_x d\psi_1 d\psi^*_1  d\psi_1' d{\psi'_1}^*\, 
  e^{-S[\psi,\psi']}
  \,, 
\end{eqnarray}
with the action 
\begin{eqnarray}\label{eq:actionpi}
  S[\psi,\psi']&=& \sum_{x,y\in\Lambda} {\psi'}^T(x) \CC \CD(x-y) \psi(y) 
  -i m\sum_{x,y\in\Lambda} {\psi'}^T(x) \CC 
  \btensor{(}{ccc} 0 & & 1 \\ 1 & &  0 \etensor{)} 
  \Gamma_5 \psi(y)\,
  \nn\\
  &=&  -\sum_n \left[ (\lambda_n+i m )(b'_n c_n+b_n c_n')+c.c.\right]\,,
\end{eqnarray}
where we have used the following 
expansion in terms of eigenfunctions of $\gamma_5 D$:
\begin{eqnarray}\label{eq:psiinef}
  \psi_1=\sum_n \left(c_n \varphi_n+ b_n \phi_n\right)\,,  
  \qquad {\rm with} 
  \qquad \gamma_5 D\varphi_n=\lambda_n \varphi_n\,,
\end{eqnarray}
and $\phi_n=\gamma_5 C^{-1} \varphi_n^*$. The above relations allow us
to show that
\begin{eqnarray}\label{eq:Z}
  Z=m^{2(n_++n_-)} 
  \left(\0{4}{a^2} +m^2\right)^{N_++N_-} \prod_{0<\lambda_n\neq 2/a} 
  (\lambda_n^2+m^2)^4\,. 
\end{eqnarray} 
with the massless limit 
$Z= \left(\0{4}{a^2}\right)^{N_++N_-} \prod_{0<\lambda_n\neq 2/a} 
\lambda_n^8$. In conclusion we find that 
\begin{eqnarray}\label{eq:Zfinal} 
Z={\rm PF}(CD)^2\,{\rm PF}(C^*D^*)^2\,.
\end{eqnarray} 
We close with a short summary. We have shown that the
construction of a theory with chirally coupled Majorana fermions on
the lattice has to deal with the usual topological obstructions
well-known from the construction of chiral fermions, even though the
total chirality is even.  The obstruction is related to the use of
chiral projection operators in the Yukawa term. This problem is resolved 
by doubling the degrees of freedom, and the Pfaffian nature of the path
integral is proven. 

\noindent {\bf Acknowledgements}

JMP would like to thank the organisers of {\it Lattice 2007} for all their 
efforts which made this inspiring conference possible. 
YI would like to thank the 
Institute of Theoretical Physics in Heidelberg for hospitality. 
We are also grateful to 
F.~Bruckmann for useful discussions. This work is supported in 
part by the Grants-in-Aid for Scientific
Research No. 17540242 from the Japan Society for the Promotion of Science.

\end{document}